\input harvmac
\input graphicx
\input color

\def\Title#1#2{\rightline{#1}\ifx\answ\bigans\nopagenumbers\pageno0\vskip1in
\else\pageno1\vskip.8in\fi \centerline{\titlefont #2}\vskip .5in}

%
%
\ifx\includegraphics\UnDeFiNeD\message{(NO graphicx.tex, FIGURES WILL BE IGNORED)}
\def\figin#1{\vskip2in}
\else\message{(FIGURES WILL BE INCLUDED)}\def\figin#1{#1}
\fi
\def\Fig#1{Fig.~\the\figno\xdef#1{Fig.~\the\figno}\global\advance\figno
 by1}
%
%
%
%

\font\ticp=cmcsc10

\def \purge#1 {\textcolor{magenta}{#1}}
\def \new#1 {\textcolor{blue}{#1}}
\def\comment#1{}

\def\\{\cr}
\def\text#1{{\rm #1}}
\def\frac#1#2{{#1\over#2}}

\def\hf{{1\over 2}} 



\def\roughly#1{\mathrel{\raise.3ex\hbox{$#1$\kern-.75em\lower1ex\hbox{$\sim$}}}}
\font\bbbi=msbm10 
\def\mathbb#1{\hbox{\bbbi #1}}

\def\mthsu{\mathsurround=0pt  }
\def\leftrightarrowfill{$\mthsu \mathord\leftarrow\mkern-6mu\cleaders
  \hbox{$\mkern-2mu \mathord- \mkern-2mu$}\hfill
  \mkern-6mu\mathord\rightarrow$}
\def\overleftrightarrow#1{\vbox{\ialign{##\crcr\leftrightarrowfill\crcr\noalign{\kern-1pt\nointerlineskip}$\hfil\displaystyle{#1}\hfil$\crcr}}}
\overfullrule=0pt

%
%
\lref\AMPS{
  A.~Almheiri, D.~Marolf, J.~Polchinski and J.~Sully,
  ``Black Holes: Complementarity or Firewalls?,''
  JHEP {\bf 1302}, 062 (2013).
  [arXiv:1207.3123 [hep-th]].
}
\lref\HaPr{
  P.~Hayden, J.~Preskill,
  ``Black holes as mirrors: Quantum information in random subsystems,''
JHEP {\bf 0709}, 120 (2007).
[arXiv:0708.4025 [hep-th]].
}
\lref\GiShtwo{
  S.~B.~Giddings and Y.~Shi,
  ``Effective field theory models for nonviolent information transfer from black holes,''
[arXiv:1310.5700 [hep-th]], Phys.\ Rev.\ D (in press).
}
\lref\NVNLT{
  S.~B.~Giddings,
  ``Modulated Hawking radiation and a nonviolent channel for information release,''
[arXiv:1401.5804 [hep-th]].
}
\lref\BHQIUE{
  S.~B.~Giddings,
  ``Black holes, quantum information, and unitary evolution,''
  Phys.\ Rev.\ D {\bf 85}, 124063 (2012).
[arXiv:1201.1037 [hep-th]].
}
\lref\SGmodels{
  S.~B.~Giddings,
   ``Models for unitary black hole disintegration,''  Phys.\ Rev.\ D {\bf 85}, 044038 (2012)
[arXiv:1108.2015 [hep-th]].
}
\lref\NLvC{
  S.~B.~Giddings,
  ``Nonlocality versus complementarity: A Conservative approach to the information problem,''
Class.\ Quant.\ Grav.\  {\bf 28}, 025002 (2011).
[arXiv:0911.3395 [hep-th]].
}
\lref\NVNL{
  S.~B.~Giddings,
  ``Nonviolent nonlocality,''
  Phys.\ Rev.\ D {\bf 88},  064023 (2013).
[arXiv:1211.7070 [hep-th]].
}
\lref\GiShone{
  S.~B.~Giddings and Y.~Shi,
  ``Quantum information transfer and models for black hole mechanics,''
Phys.\ Rev.\ D {\bf 87}, 064031 (2013).
[arXiv:1205.4732 [hep-th]].
}
\lref\Hawk{
  S.~W.~Hawking,
  ``Particle Creation By Black Holes,''
  Commun.\ Math.\ Phys.\  {\bf 43}, 199 (1975)
  [Erratum-ibid.\  {\bf 46}, 206 (1976)].
}
\lref\Braunstein{
  S.~L.~Braunstein, S.~Pirandola and K.~\.Zyczkowski,
  ``Entangled black holes as ciphers of hidden information,''
Physical Review Letters 110, {\bf 101301} (2013).
[arXiv:0907.1190 [quant-ph]].
}
\lref\miningrefs{
  W.~G.~Unruh and R.~M.~Wald
  ``How to mine energy from a black hole,''
Gen.\ Relat.\ Gravit.\ {\bf 15}, 195 (1983)\semi
  A.~E.~Lawrence and E.~J.~Martinec,
  ``Black hole evaporation along macroscopic strings,''
Phys.\ Rev.\ D {\bf 50}, 2680 (1994)
[hep-th/9312127]\semi
  V.~P.~Frolov and D.~Fursaev,
  ``Mining energy from a black hole by strings,''
Phys.\ Rev.\ D {\bf 63}, 124010 (2001)
[hep-th/0012260]\semi
  V.~P.~Frolov,
 ``Cosmic strings and energy mining from black holes,''
Int.\ J.\ Mod.\ Phys.\ A {\bf 17}, 2673 (2002).
}
\lref\Wald{
  R.~M.~Wald,
 ``Gravitation, thermodynamics, and quantum theory,''
Class.\ Quant.\ Grav.\  {\bf 16}, A177 (1999).
[gr-qc/9901033].
}
\lref\Witt{
  E.~Witten,
 ``On string theory and black holes,''
Phys.\ Rev.\ D {\bf 44}, 314 (1991).
}
\lref\Trieste{
  S.~B.~Giddings,
  ``Quantum mechanics of black holes,''
  in proceedings of High-energy physics and cosmology, Summer School, Trieste, Italy, June 13-July 29, 1994
[hep-th/9412138].
}
\lref\Math{
  S.~D.~Mathur,
  ``The Information paradox: A Pedagogical introduction,''
Class.\ Quant.\ Grav.\  {\bf 26}, 224001 (2009).
[arXiv:0909.1038 [hep-th]].
}
\lref\Braunstein{
  S.~L.~Braunstein, S.~Pirandola and K.~\.Zyczkowski,
  ``Entangled black holes as ciphers of hidden information,''
Physical Review Letters 110, {\bf 101301} (2013).
[arXiv:0907.1190 [quant-ph]].
}
\lref\Page{
  D.~N.~Page,
  ``Particle Emission Rates from a Black Hole: Massless Particles from an Uncharged, Nonrotating Hole,''
Phys.\ Rev.\ D {\bf 13}, 198 (1976).
}
\lref\Hartle{
  J.~B.~Hartle,
  {\sl Gravity: An introduction to Einstein's general relativity,}
San Francisco, USA: Addison-Wesley (2003) 582 p.
}
\lref\CGHS{
  C.~G.~Callan, Jr., S.~B.~Giddings, J.~A.~Harvey and A.~Strominger,
  ``Evanescent black holes,''
Phys.\ Rev.\ D {\bf 45}, 1005 (1992).
[hep-th/9111056].
}
\lref\ChFu{
  S.~M.~Christensen and S.~A.~Fulling,
  ``Trace Anomalies and the Hawking Effect,''
Phys.\ Rev.\ D {\bf 15}, 2088 (1977).
}
\lref\HaHa{
  J.~B.~Hartle and S.~W.~Hawking,
  ``Path Integral Derivation of Black Hole Radiance,''
Phys.\ Rev.\ D {\bf 13}, 2188 (1976).
}
\lref\Hawkunc{
  S.~W.~Hawking,
  ``Breakdown of Predictability in Gravitational Collapse,''
Phys.\ Rev.\ D {\bf 14}, 2460 (1976).
}
\lref\GiddingsSJ{
  S.~B.~Giddings,
  ``Black hole information, unitarity, and nonlocality,''
Phys.\ Rev.\ D {\bf 74}, 106005 (2006).
[hep-th/0605196].
}
\lref\GiShone{
  S.~B.~Giddings and Y.~Shi,
  ``Quantum information transfer and models for black hole mechanics,''
Phys.\ Rev.\ D {\bf 87}, no. 6, 064031 (2013).
[arXiv:1205.4732 [hep-th]].
}
\lref\NVNL{
  S.~B.~Giddings,
  ``Nonviolent nonlocality,''
Phys.\ Rev.\ D {\bf 88}, 064023 (2013).
[arXiv:1211.7070 [hep-th]].
}
\lref\NVNLFT{
  S.~B.~Giddings,
  ``Nonviolent information transfer from black holes: A field theory parametrization,''
Phys.\ Rev.\ D {\bf 88}, no. 2, 024018 (2013).
[arXiv:1302.2613 [hep-th]].
}
\lref\GiShtwo{
  S.~B.~Giddings and Y.~Shi,
  ``Effective field theory models for nonviolent information transfer from black holes,''
Phys.\ Rev.\ D {\bf 89}, no. 12, 124032 (2014).
[arXiv:1310.5700 [hep-th]].
}
\lref\NVNLT{
  S.~B.~Giddings,
  ``Modulated Hawking radiation and a nonviolent channel for information release,''
Phys.\ Lett.\ B {\bf 738}, 92 (2014).
[arXiv:1401.5804 [hep-th]].
}
\lref\HaPr{
  P.~Hayden, J.~Preskill,
  ``Black holes as mirrors: Quantum information in random subsystems,''
JHEP {\bf 0709}, 120 (2007).
[arXiv:0708.4025 [hep-th]].
}
\lref\Susstrans{
  L.~Susskind,
  ``The Transfer of Entanglement: The Case for Firewalls,''
[arXiv:1210.2098 [hep-th]].
}
\lref\Unru{
  W.~G.~Unruh,
  ``Notes on black hole evaporation,''
Phys.\ Rev.\ D {\bf 14}, 870 (1976).
}
\lref\GiddingsBE{
  S.~B.~Giddings,
  ``(Non)perturbative gravity, nonlocality, and nice slices,''
Phys.\ Rev.\ D {\bf 74}, 106009 (2006).
[hep-th/0606146].
}
\lref\Unruorig{
  W.~G.~Unruh,
  ``Origin of the Particles in Black Hole Evaporation,''
Phys.\ Rev.\ D {\bf 15}, 365 (1977).
}
\lref\EGK{
  M.~Eune, Y.~Gim and W.~Kim,
  ``Something special at the event horizon,''
Mod.\ Phys.\ Lett.\ A {\bf 29}, no. 40, 1450215 (2014).
[arXiv:1401.3501 [hep-th]].
}
\lref\GiKi{
  Y.~Gim and W.~Kim,
  ``A Quantal Tolman Temperature,''
Eur.\ Phys.\ J.\ C {\bf 75}, no. 11, 549 (2015).
[arXiv:1508.00312 [gr-qc]].
}
\lref\NSW{
  Y.~Nomura, F.~Sanches and S.~J.~Weinberg,
  ``Relativeness in Quantum Gravity: Limitations and Frame Dependence of Semiclassical Descriptions,''
JHEP {\bf 1504}, 158 (2015).
[arXiv:1412.7538 [hep-th]];
  ``Black Hole Interior in Quantum Gravity,''
Phys.\ Rev.\ Lett.\  {\bf 114}, 201301 (2015).
[arXiv:1412.7539 [hep-th]].
}
\Title{
\vbox{\baselineskip12pt  
}}
{\vbox{\centerline{Hawking radiation, the Stefan-Boltzmann law,} \centerline{and unitarization}
}}

\centerline{{\ticp 
Steven B. Giddings\footnote{$^\ast$}{Email address: giddings@physics.ucsb.edu}
} }
\centerline{\sl Department of Physics}
\centerline{\sl University of California}
\centerline{\sl Santa Barbara, CA 93106}
\vskip.10in
\centerline{\bf Abstract}
Where does Hawking radiation originate?  A common picture is that it arises from excitations very near or at the horizon, and this viewpoint has supported the ``firewall" argument and arguments for a key role for the UV-dependent entanglement entropy in describing the quantum mechanics of black holes.  However, closer investigation of  both the total emission rate and the stress tensor of Hawking radiation  supports the statement that its source is
a near-horizon quantum region, or ``atmosphere," whose radial extent is set by the horizon radius scale.  This is potentially important, since Hawking radiation needs to be modified to restore unitarity, and a natural assumption is that the scales relevant to such modifications are comparable to those governing the Hawking radiation.  Moreover, related discussion suggests a resolution to questions regarding extra energy flux in ``nonviolent" scenarios, that does not spoil black hole thermodynamics as governed by the Bekenstein-Hawking entropy.

\vskip.3in
\Date{}

Hawking radiation is commonly perceived as originating from the horizon of a black hole.  One reason for this is the structure of Hawking's original calculation\refs{\Hawk}:  highly blueshifted modes just outside the horizon, which are entangled with similar inside excitations, can be described as evolving to become the radiation.  This view is buttressed by a nice match to the thermal description of the observations of detectors at constant radius $r$.  These detectors are properly accelerating, and so experience the Unruh effect with a temperature that is related to Hawking's by a blueshift, in accord with the Tolman law; see for example \refs{\Wald}.

It is important, however, to check this picture, since the requirement of unitarity of the black hole decay tells us that the Hawking radiation must be modified.  If we wish to understand what kind of modification is needed, and where it occurs, we should first fully understand the properties of the Hawking radiation, which is responsible for the problem of information loss to begin with.  This is emphasized, for example, by the structure of the ``firewall" argument:  if one presupposes a near-horizon origin of the Hawking radiation, and that the corresponding near-horizon excitations must therefore be modified  in order to restore unitarity, one concludes that the state is very singular, with an enormous energy density also rendering the spacetime geometry singular at the 
horizon\refs{\Trieste\Math\Braunstein-\AMPS}.

So, in order to better understand both where unitarizing modifications might appear, and also other aspects of the thermodynamics of black holes, we seek other tests for the source of the Hawking radiation.   

One way to infer the size of a radiating body is via the Stefan-Boltzmann law, giving the radiated power (in the case of two polarization degrees of freedom, {\it e.g.} photons)
\eqn\SB{{dE\over dt} = \sigma_{S} A T^4}
in terms of the area $A$ of an emitting black body, and its temperature; here $\sigma_{S}=\pi^2/60$ is the Stefan-Boltzmann constant.  From this, one finds the area of the emitting surface from the power and the temperature, which for Hawking radiation we expect to be the Hawking temperature.  
A complication, however, is that a black hole emits as a gray body -- it is not precisely thermal.  But, once gray-body factors are taken into account, numerical calculation\refs{\Page} shows that the emission rate {\it exceeds} the rate \SB\ for particles with spin $\leq1$ if $A$ is taken to be the horizon area -- suggesting a larger effective emitting surface.  Specifically, considering for example photon emission, ref.~\Page\ (see eq.~(29) and below) shows a total rate for a black hole of mass $M$
\eqn\pagerate{ {dE\over dt}= 3.4 \times 10^{-5} M^{-2}\ ,}
 as compared to a rate 
 \eqn\horrate{{dE\over dt}= 2.1 \times 10^{-5} M^{-2}}
 from \SB\ if $T=1/(8\pi M)$ is the Hawking temperature and $A=16\pi M^2$ the horizon area.

The calculation and conclusions can be sharpened by looking at the power spectrum, which can be expressed as
\eqn\pspec{{dE\over dt d\omega} = {1\over \pi}\sum_l (2l+1) \omega {\Gamma_{\omega l}\over e^{\beta \omega} -1} = {1\over \pi^2} {\omega^3\over e^{\beta \omega} -1} \sigma(\omega)}
for two degrees of freedom, where $l$ is the orbital angular momentum, $\Gamma_{\omega l}$ are the gray-body factors, and $\beta=1/T$.  In the second equality, the spectrum has been related to the absorption cross section at frequency $\omega$,
\eqn\absc{\sigma(\omega) =  {\pi\over \omega^2} \sum_l (2l+1)\Gamma_{\omega l}\ .}
For a spherical blackbody of area $A=4\pi r^2$, $\sigma(\omega)=\pi r^2 = A/4$, and \SB\ is reproduced.  In the case of Hawking radiation, the gray-body factors vary nontrivially with $\omega$, but in the large-$\omega$ limit, 
\eqn\HEsigma{\sigma(\omega)\rightarrow \pi R_a^2}
where
\eqn\RAdef{R_a = 3\sqrt3 M = {3\sqrt3\over 2} R\ }
and $R=2M$ is the Schwarzschild radius.
This limit is the geometric-optics, massless limit, and so this result can be understood from the effective potential (see {\it e.g.} \refs{\Hartle}) for a classical massless particle.  Here absorption is perfect for $l<\omega R_a$, and vanishes for $l>\omega R_a$, and so
\eqn\gammaapprox{\Gamma_{\omega l}\approx \theta(\omega R_a - l)\ ,}
giving \HEsigma, and yielding a high-energy power spectrum \pspec\ matching that of \Page.

Thus  the effective emitting area for the Hawking radiation can be read off from this high-energy emission, and is $A=4\pi R_a^2$; the effective emitting radius $R_a$ is considerably outside the horizon radius, which is indicative of a source well outside the horizon.
Note that for lower-energy modes, where quantum effects become more relevant, the gray-body factors are suppressed from unity.  Since most of the emission is in such modes, this yields\Page\ a total power \pagerate\ that is suppressed from \SB\ evaluated with $A=4\pi R_a^2$.

Since the statement that the source of the Hawking radiation is well outside the horizon runs contrary to various perceptions, we should try to test it by other means.  A more refined picture of the Hawking radiation comes from examining its stress tensor.  This is particularly tractable in the case of a two-dimensional metric, taken to be of the form
\eqn\tdmet{ds^2=-f(r)dt^2+{dr^2\over f(r)} = f(r)(-dt^2+dx^2)=-f(r)dx^+dx^-}
where 
\eqn\tortdef{{dx} = {dr\over f(r)}\ .}
and $x^\pm=t\pm x$.  
The conformal coordinate $x$ is sometimes referred to as a tortoise coordinate.  For the two-dimensional black hole of \Witt, studied in the soluble collapse models of \refs{\CGHS}, 
\eqn\ftd{f(r) = 1-e^{-2(r-R)}\ .}
However, the metric \tdmet\ may also be thought of the metric induced on a cosmic string that threads a higher-dimensional black hole, allowing us to probe that case as well.

The expectation value of the stress tensor for Hawking radiation can be computed via the conformal anomaly\refs{\ChFu,\CGHS}:
\eqn\stress{\eqalign{\vev{T_{--}}&={1\over 24\pi}\left[{\partial_-^2 f\over f} -{3\over 2}{(\partial_- f)^2\over f^2}\right]+t_-(x^-)\cr
\vev{T_{++}}&={1\over 24\pi}\left[{\partial_+^2 f\over f} -{3\over 2}{(\partial_+ f)^2\over f^2}\right]+t_+(x^+)\cr
\vev{T_{+-}}&= -{1\over 24\pi}\left({\partial_+\partial_- f\over f} -{\partial_+f\partial_-f\over f^2}\right)
}}
where $t_-(x^-)$ and $t_+(x^+)$ are arbitrary functions characterizing the particular state.  It is readily verified that \stress\ is conserved.  Indeed, the conformal anomaly determines $\vev{T_{+-}}$, and then conservation fixes $\vev {T_{--}}$ and $\vev{T_{++}}$, up to the functions $t_\pm$.

Eq.~\stress\ may be written in terms of $r$-derivatives of $f$, denoted by primes, using \tortdef.  This gives
\eqn\fstress{\eqalign{\vev{T_{--}}&={1\over 96\pi}\left[f f^{\prime\prime}-\hf (f')^2\right] +t_-\cr
\vev{T_{++}}&={1\over 96\pi}\left[f f^{\prime\prime}-\hf (f')^2\right] +t_+\cr
\vev{T_{+-}}&={1\over 96\pi} f f^{\prime\prime}\ .
}}
For the Hartle-Hawking\refs{\HaHa} or Unruh\refs{\Unru} states, regularity of $\vev{T_{\mu\nu}}$ at the future horizon, checked in terms of the {\it Kruskal} components of $\vev{T_{\mu\nu}}$, then implies 
\eqn\outt{t_-={1\over 192\pi} [f'(R)]^2\ .}
Since the other terms in $\vev{T_{--}}$ vanish asymptotically at $r\rightarrow\infty$, $t_-$ is the asymptotic Hawking flux.  For the Hartle-Hawking vacuum, this flux is balanced by incoming flux, $t_+=t_-$, and $\vev{ T_{\mu\nu}}$ is also regular on the past horizon. For the Unruh vacuum, $t_+=0$, so there is no incoming asymptotic flux, but there is a negative energy flux into the horizon.  Note that $\vev{T_{--}}$ also vanishes to {\it next} order in $r-R$, as can be readily verified by taking its $r$-derivative, from \fstress; that is, $\vev{T_{--}}$ vanishes as $f^2(r)$ at $r=R$.

We now see properties that support the preceding claim.  The outward Hawking flux $\vev{T_{--}}$ can be converted into that in an orthonormal frame ({\it c.f.} \tdmet) by multiplying by $1/f$, but the resulting proper $\vev{T_{\hat -\hat -}}$ still vanishes at the horizon; the proper outward flux builds up from there, over a range of $r\sim R$, to its asymptotic value.
That is, the outgoing Hawking flux, as measured by its stress tensor, originates not at the horizon, but in a larger quantum region or atmosphere.  
For the Hartle-Hawking vacuum, $\vev{T_{\hat0\hat1}}$ identically vanishes due to cancellation between ingoing and outgoing flux. 
For the Unruh vacuum, $\vev{T_{\hat0\hat1}}$ is nonvanishing at the horizon  due to the negative {\it influx}\refs{\Unruorig}\foot{Indeed, following the first appearance of this paper, the author became aware of  \Unruorig\ which gave closely related arguments, for an origin of Hawking particles in the vicinity of a black hole rather than from the collapsing body that formed it. Unruh's arguments were based on 1) the fact that energy appears outside the black hole and is compensated by the negative influx; 2) the failure of infalling observers to detect particles near the horizon (see also \refs{\GiddingsSJ}); and 3) the existence of stimulated emission due to an emitter falling into a black hole.  Refs.~\refs{\EGK,\GiKi} have also investigated the role of the negative energy density at the horizon, and pointed out vanishing of an effective ``Tolman" temperature there, and refs.~{\NSW} makes possibly related comments about negative influx.}    of energy described by $\langle T_{++}\rangle$.  This energy flux at a near-horizon coordinate $r$ does satisfy a two-dimensional version of the Stefan-Boltzmann law of the form
\eqn\tdsb{{dE\over dt} = -\vev{T_{\hat0\hat1}} = \sigma_2 T^2(r)\ ,}
 where $T(r)$ is the locally blueshifted temperature, which is seen by the locally accelerated observers at constant $r$, and $\sigma_2$ is a constant.  But this flux does not originate from the outward-going Hawking particles.   Over the quantum region outside the horizon spanning a range $\Delta r\sim R$, the negative flux $\vev{T_{++}}$ transitions to the nonzero positive $\vev{T_{--}}$.  

While these statements are made in two dimensions, they are directly pertinent to higher-dimensional black holes.  Specifically, the quantum atmosphere of a higher-dimensional black hole may be probed by threading the black hole with a cosmic string.  Then, any modes along the string provide a direct channel for Hawking emission that avoids the usual angular momentum barriers.  This means such a black hole emits more quickly; this gives a simple example of the process of mining a black hole\refs{\miningrefs}.  The $1+1$-dimensional metric on the string is induced from that of the ambient spacetime, so for $D$-dimensional Schwarzschild,
\eqn\schwf{f(r)= 1-\left({R\over r}\right)^{D-3}\ .}
Here, too, the outward Hawking flux builds up over a range  $\Delta r\sim R$ outside the horizon.

We have thus found two arguments that the Hawking radiation originates from a range of $r$ comparable to $R$ outside the horizon of a Schwarzschild black hole -- which may be referred to as the atmosphere --  and not from a small region at the black hole horizon.

It is also informative to look at the wavelength of the radiation, which for a typical Hawking quantum takes the value $\lambda\approx \lambda_T$ where
\eqn\thermw{\lambda_T = {2\pi\over T} = 8\pi^2 R \approx 79\, R\ .}
Thus the horizon size is much smaller than the thermal wavelength, in contrast to typical discussions of blackbody radiation.  
One can also examine the wavelength of the near-horizon blueshifted modes whose occupation ultimately yields the Hawking radiation.  If we consider such a mode centered at some near-horizon $r$, with the typical wavepacket width $\Delta x=\lambda_T$, then the near-horizon limit of \tortdef\ shows that the wavepacket edges will be at 
\eqn\waveends{(r-R)_{\rm edges} \approx (r-R) e^{\pm f'(R)\lambda_T/2} =(r-R)e^{\pm 4\pi^2}\ }
where the last equality uses $f(r)$ for four-dimensional Schwarzschild.  
So such typical modes span a range of $r$ much larger than the separation of their centers from the horizon.  This is in accord with the observations of \GiddingsBE, that the modes do not separate from their ``Hawking partners" inside the black hole until after they separate from the vicinity of the black hole itself.
So any discussion of observations of near-horizon observers, at scales small as compared to the separation from the horizon, $\delta r\ll r-R$, or any such manipulation of these modes\AMPS, involves trying to describe these modes at a scale that is well within their typical, thermal, wavelength.   

While this black hole story does depart from the usual black body situation, it certainly {\it doesn't} appear to support a shorter distance origin of the Hawking radiation.  The combined observations of this paper instead suggest that the ultraplanckian origin of the Hawking excitations seen in \Hawk\ is very much an artifact of that particular way to calculate the Hawking effect.

The statement of a longer-distance origin for Hawking radiation has potentially important implications for the question of unitarization.  Hawking radiation is responsible for a loss of unitarity\refs{\Hawkunc}.  This tells us that the Hawking state must be modified to save quantum mechanics.  If the Hawking radiation originates from the atmosphere of a black hole, and not from the horizon, it is reasonable and natural to expect that the new effects unitarizing it are also operational there, rather than right at the horizon.  This is exactly what is proposed in the simplest ``nonviolent" scenarios of \refs{\GiddingsSJ,\NLvC\SGmodels\BHQIUE\GiShone\NVNL\NVNLFT\GiShtwo-\NVNLT} -- in contrast to the firewall picture advocated by \refs{\AMPS}.  

In fact, the preceding observations suggest a way to approach one of the questions asked about the nonviolent approach.  If one describes this approach from the viewpoint of an effective theory approximation,  it involves extra couplings, beyond those of semiclassical gravity, to excitations in the atmosphere of a black hole.  These couplings depend on the quantum state of the black hole, and ultimately are responsible the entanglement transfer from the black hole to its environment\refs{\HaPr,\GiShone,\Susstrans} which is need to unitarize black hole decay.  As pointed out in \refs{\SGmodels,\BHQIUE,\NVNL\NVNLFT-\GiShtwo}, such couplings typically also produce extra energy flux from the black hole, due to the extra channels they introduce.  While models based on effective couplings to the stress tensor might for example minimize this effect\NVNLT, it is still a challenge to avoid extra flux.  

However, such extra energy flux, if carried via modes with typical energies $\sim T$, is not necessarily in contradiction with black hole thermodynamics.  This is because, as we have noted, the Hawking power is significantly below that of a black body, and this means additional flux is  possible, respecting \SB, without modifying the temperature $T$.  This can be thought of as arising due to additional couplings between black hole excitations with typical energies $\sim T$ and the modes of the black hole environment, that make up for some of the suppression in the low-frequency gray-body factors.   (Note\Page\ that this suppression is particularly strong for gravitons.)  In particular, couplings via the stress tensor as in \refs{\NVNLT} are expected to alter these factors at low energies.
Thus additional couplings of the black hole to exterior,  like those described in the effective field theory approximation to NVNL,  can modify the energy flux, without modifying the Bekenstein-Hawking formula for the entropy of a black hole.

Such couplings could also increase the absorption of modes with frequencies $\omega\sim R$, and this is suggested by \pspec.  Indeed, if the black hole were sustained by a thermal flux in an equilibrium configuration, the increase in the emitted energy would need to be balanced by an increase in the absorbed energy.  However, such couplings would {\it not} need to significantly affect modes at higher frequencies, $\omega\gg1/R$, and so can approximately respect the equivalence principle for such modes.

In conclusion, this note has presented evidence that the source of Hawking radiation is  a quantum region of size $\Delta r\sim R$ outside the black hole horizon.  If the new effects modifying local quantum field theory, which are required to unitarize Hawking decay, have the same characteristic scales as the Hawking radiation -- in other words, if the solution has the same scales as the problem -- these couplings would also be expected to extend to  radii $r\sim3\sqrt{3}R/2 $ or larger, matching the scale size proposed in the simplest non-violent nonlocality scenarios\refs{\GiddingsSJ,\NLvC\SGmodels\BHQIUE\GiShone\NVNL\NVNLFT\GiShtwo-\NVNLT}.  In order to maintain a thermodynamic description with temperature $T$, these couplings should also be primarily be to modes with energies $\omega\sim T$.  Thus, these simple arguments  strongly suggest the scales relevant to a theory unitarizing quantum evolution of black holes.

\bigskip\bigskip\centerline{{\bf Acknowledgments}}\nobreak

This work  was supported in part by the Department of Energy under Contract {DE-SC}0011702.  The author thanks J. Hartle for discussions, and D. Page for an email message clarifying the results of \Page.  He also thanks D. Marolf for confirming that some believe Hawking radiation originates at the horizon.

\listrefs
\end